\documentclass[12pt]{iopart}
\usepackage{graphicx}
\usepackage{color}

\begin{document}

\title[Self-consistent calculation of the reactor antineutrino spectra including forbidden transitions] {Self-consistent calculation of the reactor antineutrino spectra including forbidden transitions}

\author{J. Petkovi\'{c}}
\address{Department of Physics, Faculty of Science, University of Zagreb, 10000 Zagreb, Croatia}
\author{T. Marketin}
\address{Ericsson Nikola Tesla d.d., Krapinska 45, 10000 Zagreb, Croatia}
\author{G. Mart\'{i}nez-Pinedo}
\address{GSI Helmholtzzentrum f\"{u}r Schwerioneneforschung, Planckstra{\ss}e 1, 64291 Darmstadt, Germany}
\address{Institut f\"{u}r Kernphysik (Theoriezentrum), Technische Universit\"{a}t Darmstadt, 64289 Darmstadt, Germany}
\author{N. Paar}
\address{Department of Physics, Faculty of Science, University of Zagreb, 10000 Zagreb, Croatia}

\ead{npaar@phy.hr}
\vspace{10pt}
\begin{indented}
\item[]March 2019
\end{indented}

\begin{abstract}
With the goal of determining the $\theta_{13}$ neutrino oscillation mixing angle, the measurements of reactor antineutrino fluxes at the Double Chooz, RENO and Daya Bay experimental facilities have uncovered a systematic discrepancy between the number of observed events and theoretical expectations. In the \emph{ab initio} approach, the total reactor antineutrino spectrum is a weighted sum of spectra resulting from all $\beta$ branches of all fission products in the reactor core. At all three facilities a systematic deviation of the number of observed events from the number of predicted events was noticed, i.e., approximately 6\% of the predicted neutrinos were not observed. This discrepancy was named the reactor neutrino anomaly. In theoretical studies it is assumed that all the decays are allowed in shape, but a quarter of all transitions are actually forbidden and may have a complex energy dependence that will affect the total reactor antineutrino spectrum. In order to estimate the effect of forbidden transitions, we perform a fully self-consistent calculation of spectra from all contributing transitions and compare the results with a purely allowed approximation.
\end{abstract}

%
%
%
%
%

\section{Introduction}
Nuclear reactors are the most intense man-made sources of antineutrinos, and they were involved in neutrino physics from their first detection. Since then, they have been a key component in the study of neutrino properties - most importantly the study of neutrino oscillations. In particular, the experimental effort was focused on the determination of neutrino mixing angles $\theta_{13}$ and $\theta_{23}$, and more recently on the precise measurement of the final angle $\theta_{13}$. In these measurements information on the reactor antineutrino flux and spectrum is integral to the analysis, and affects the final values. 

In the quest for the precise determination of the $\theta_{13}$ mixing angle, the Double Chooz~\cite{Abe2012}, RENO~\cite{Ahn2012} and Daya Bay~\cite{An2012} experimental facilities have provided a wealth of information. During the analysis of the data, it was noticed that the measured antineutrino spectrum was systematically lower than was predicted, in all three facilities. The question was further complicated by a reevaluation of the reactor flux, which produced a total discrepancy between the measured and the expected antineutrino spectrum of approximately 6\%~\cite{Mention2011,Huber2011} - the so-called ``reactor antineutrino anomaly''\cite{Hayes2016}. Further reports have confirmed the existence of the anomaly and uncovered an unexplained structural feature in the antineutrino spectrum at energies between 5 MeV and 7 MeV of antineutrino energy~\cite{An2015,An2016,Dwyer2015,Dwyer2015a}. The existence of the anomaly has spurred an active discussion on the nature of neutrinos and the possible existence of sterile neutrinos. But the uncertainties of the determination of theoretical lepton spectra are relatively large and may explain the anomaly, especially in view that no anomalous neutrino disappearance was observed in a recent report~\cite{Aartsen2016}.

Theoretical determination of the antineutrino spectra is based on available data, i.e the energies of the transitions, their parity and angular momentum, and corresponding branching ratios. Two approaches are used in order to determine the total reactor lepton spectra: (i) the conversion method, and (ii) the ``ab initio'' summation method. With the conversion method one uses the precisely measured aggregate electron spectrum to fit a relatively small (compared to the total number of measured transitions) set of virtual transitions from which one obtains the corresponding antineutrino spectrum. Measurements were performed in Grenoble on $^{235}$U, $^{239}$Pu and $^{241}$Pu~\cite{Feilitzsch1982,Hahn1989,Schreckenbach1985} and Garching on $^{238}$U~\cite{Haag2014}. This method requires no information on, but also provides no insight into, fission yields and branching ratios. 

The ``ab initio'' summation method takes the opposite approach. By combining individual electron and antineutrino spectra for each branching ratio of each fission fragment, where the endpoint energies and relative probabilities are taken from data, the total lepton spectra are obtained. While this method should, in principle, be able to reproduce the spectra in full, the results are dependent on the accuracy of the available data. In particular, the incorrect assignment of feeding probabilities of excited states in the daughter nuclei can have a significant impact~\cite{Algora2010,Fallot2012}. 

While approximately 25\% of all transitions are forbidden, in the limit of a vanishing electron mass, the shape factors for forbidden transitions are symmetric under the exchange of electron and antineutrino energies~\cite{Huber2011}. Thus, in most studies the allowed shape factor is also used for forbidden transitions. However, a recent study reveals significant differences in the total antineutrino spectra depending on the treatment of forbidden transitions~\cite{Hayes2014}. The authors have treated all unique forbidden transitions as unique first-forbidden transitions, and all nonunique forbidden transitions as either allowed, unique first-forbidden $2^{-}$ transitions, nonunique $0^{-}$ or nonunique $1^{-}$ transitions. Treating all transitions equally introduces a systematic uncertainty in the results, but the authors have shown a consistent and non-negligible effect of forbidden transitions on the total antineutrino spectra coming from products of fissile material in a typical reactor. {More recently, in Ref.~\cite{Hayen2019} microscopic calculations of the first-forbidden transitions in 
reactor antineutrino spectra have been performed. Explicit calculation of the shape factor showed differences in cumulative 
electron and antineutrino spectra in comparison to usually employed approximations. It has been shown that forbidden decays represent an essential ingredient for reliable understanding of reactor antineutrino spectra~\cite{Hayen2019}, thus further research to assess 
their role is called for.}

Very recently, a study performed at the Daya Bay experimental facility observed a correlation between reactor core fuel evolution and changes in the reactor antineutrino flux and its energy spectrum. In fact, a careful analysis of 2.2 million inverse beta decay events over 1230 days, a discrepancy between the assumed and measured effective fission fractions was uncovered with major impact on the total antineutrino spectrum - a 7.8 \% difference for $^{235}$U. Whether this change may account for the complete reactor antineutrino anomaly, or just a significant part will require further study. 

In light of these developments, even in the case of no anomaly, it is essential to quantitatively determine the effect of first-forbidden transitions on the total lepton spectra coming from the $\beta$-decay of fission fragments of main reactor fuel materials. In this contribution we present the first fully theoretical calculation of electron and antineutrino spectra which properly takes into account the specific shape factors for each transition. We employ the relativistic Hartree-Bogoliubov (RHB) model to describe the nuclear ground state~\cite{Vretenar2005a}, and use the proton-neutron relativistic quasiparticle random phase approximation (pn-RQRPA) formulated in the canonical single-nucleon basis of the RHB model to obtain the excited states~\cite{Paar2004}. With this fully self-consistent model we have calculated the total decay rates and branching ratios for all $\beta$-unstable fission fragments, properly taking into account the shape factors of first-forbidden transitions. We generated the lepton spectra for each transition, weighted them with their corresponding branching ratios and fission yields and obtained the total electron and antineutrino spectra per fission. This process was performed for all four major contributors to the reactor antineutrino spectrum: $^{235}$U, $^{238}$U, $^{239}$Pu and $^{241}$Pu. Here we present the results for $^{235}$U and $^{239}$Pu, which are the two isotopes that provide the dominant contribution over the fuel cycle. The results for the remaining two isotopes are very similar and do not affect the final conclusions. 

\section{Evaluation of $e^{-}$ and $\bar{\nu}_{e}$ spectra} \label{sec:theory}

During the operation of a nuclear reactor the fissile isotopes fission, generating a distribution of fission products. Many of the fission products are unstable and $\beta$-decay towards stability,
\begin{equation} \label{eq:decay}
{}^{A}_{Z}X_{N} \to \, {}^{A}_{Z+1}Y_{N-1} + e^{-} + \bar{\nu}_{e}
\end{equation}
emitting an electron and an antineutrino in the process with their maximal energies begin determined by the difference in energies between the initial and final states. Of all the material in the reactor core, isotopes $^{235}$U, $^{238}$U, $^{239}$Pu and $^{241}$Pu are responsible for more than 99.7\% of all antineutrinos~\cite{An2016a}. Of those, $^{235}$U and $^{239}$Pu contribute more than 90\% of fissions in a reactor, with $^{235}$U dominating at the start of the burn-up cycle, but contributing roughly equally to $^{239}$Pu after 20000 MWD/TU.
For a particular transition between the ground state of the parent nucleus and a state (ground state or excited) in the daughter nucleus, the electron spectrum is of the form
\begin{equation} \label{eq:spec4}
S_{f}^{i}(E) = W\sqrt{W^{2} - 1}(W_{0} - W)^{2} C(W) \delta(Z,W),
\end{equation}
where $W$ and $W_{0}$ are the electron energy and the maximum electron energy, respectively, both in units of electron mass. $C(W)$ is the shape factor and $\delta(Z,W)$ is the correction factor that compensates for various approximations. The correction factor takes into account the fact that the electron is moving in the Coulomb field of the nucleus (the Fermi function), the effects of the finite size of the charge distribution $L_{0}(Z,W)$ and the weak-interaction finite-size correction $C(Z,W)$. For the treatment of these corrections we follow Ref.~\cite{Huber2011}, and neglect the corrections arising from weak magnetism and screening effects as the size of these corrections is less than 2\% for all antineutrino energies.

The shape factor is a critical quantity that determines the electron spectrum. For allowed decays it is simply equal to the Gamow-Teller strength and does not depend on energy, and thus does not affect the spectrum shape. In the case of first-forbidden transitions, the shape factor is energy dependent and can be written as
\begin{equation} \label{eq:shape}
C(W) = k\left(1 + aW + bW^{-1} + cW^{2} \right),
\end{equation}
where $k$, $ka$, $kb$ and $kc$ are given by combinations of transition matrix elements~\cite{Behrens1971,Behrens1982,Marketin2016}. Thus the electron spectrum coming from forbidden transitions can significantly differ from the shape of the allowed transitions, depending on which matrix elements dominate for a particular transition. 

\begin{figure}[htb]
  \centering
  \includegraphics[width=\linewidth]{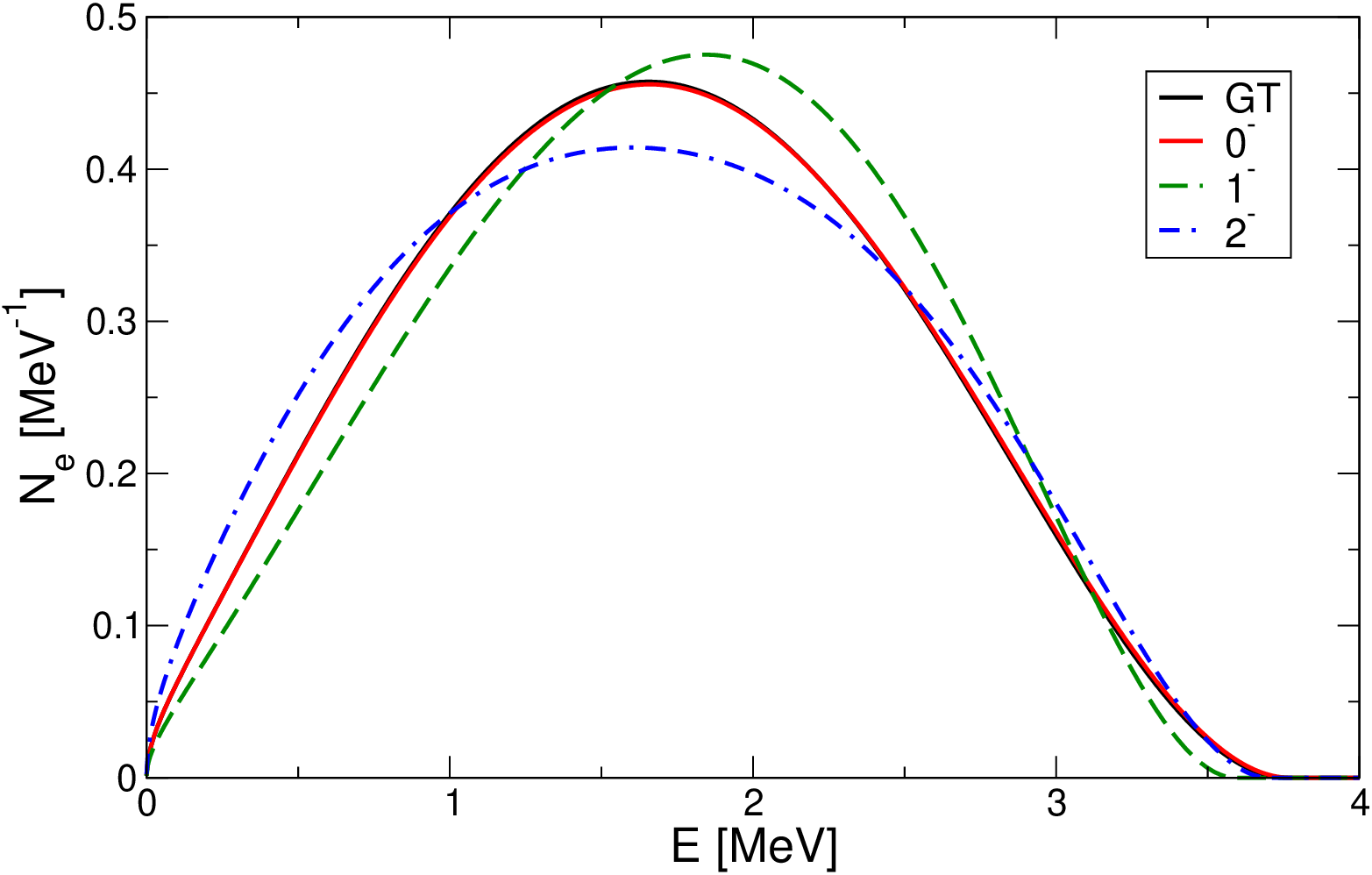}
  \caption{(color online) \label{fig:multipoles}The electron spectra for transitions of equal energy but different angular momenta. {The calculated spectra obtained with Gamow-Teller (GT) transitions (solid black line), $0^{-}$ (solid red line), $1^{-}$ (dashed green line), and  $2^{-}$ (dot-dashed blue line) transitions are separately displayed.}  The spectra are shown without any corrections.}
\end{figure}

In Fig. \ref{fig:multipoles} we show the {calculated electron spectra for a set of} hypothetical transitions of equal energy, but different angular momenta. The full black line denotes the spectrum assuming an allowed shape factor, {including only Gamow-Teller transitions}. The spectrum obtained by assuming a $0^{-}$ transitions is denoted with a full red line which follows the allowed spectrum almost completely. The shape factor of $0^{-}$ transitions consists of two terms, one of which is energy independent and dominates the total transition strength. Thus, the shape of the spectrum for $0^{-}$ transitions is almost identical to the shape of the Gamow-Teller transitions. This explains the excellent agreement that was achieved in Ref.~\cite{Sonzogni2015}, where the $\beta$ spectra for $^{92}$Rb and $^{96}$Y were described assuming allowed shape, even though both decays are dominantly ($> 95\%$) $0^{-} \to 0^{+}$. This is an important point because some of the most contributing nuclei decay by $0^{-}$ transitions (see Tables II and III in Ref.~\cite{Sonzogni2015}).

Higher angular momentum transitions have a significantly different shape from the allowed spectrum, and may have a noticeable impact on the total antineutrino spectra. In particular, the components of a $1^{-}$ transitions are found in all terms of Eq. (\ref{eq:shape}), and the relative importance of individual matrix elements will form the final transition spectrum. Typically though, the maximum of the $1^{-}$ transitions is shifted to higher energies compared to Gamow-Teller transitions, and the spectrum is slightly narrower, in total. $2^{-}$ spectra are, in general, wider than the allowed for the same transition energy. 

In many decays, first-forbidden transitions provide an appreciable contribution to the total decay rate, and thus to the total $\beta$ and $\bar{\nu}$ spectra~\cite{Moeller2003,Marketin2016,Mustonen2016}. This is particularly true in very neutron-rich nuclei where additional neutrons in the higher shell enable more parity-changing transitions. As the most neutron-rich fission products decay towards stability, on average they decay 6 times. In this way, the distortion of the lepton spectra arising from forbidden transitions may accumulate to produce a noticeable effect on the total observed spectrum. To establish a quantitative measure of this effect, we have determined the electron and antineutrino spectra for the four contributing isotopes in reactors using a fully theoretical ``ab initio'' approach. Using the decay data obtained in a large scale calculation of $\beta$-decay properties of r-process nuclei~\cite{Marketin2016}, we have generated the $\beta$ and $\bar{\nu}$ spectra for each transition $S_{f}^{i}$. To obtain the lepton spectrum arising from a single decay of a particular nuclide we weigh transition spectra $S_{f}^{i}$ with their respective branching ratios and sum,
\begin{equation} \label{eq:spec3}
S_{f}(E) = \sum_{i} \frac{{\lambda}_{i}}{{\lambda}_{tot}} S_{f}^{i}(Z,A,E_{max},E,J^{\pi}).
\end{equation}
Here $f$ denotes a particular fission fragment, and $i$ denotes a transition to a particular final state in the daughter nucleus, where the sum runs over all energetically allowed transitions. Finally, these spectra are weighted by their respective cumulative fission yields and summed in order to obtain the total electron and antineutrino spectra for a particular actinide. 
\begin{equation} \label{eq:spec2}
S_{k}(E) = \sum_{f} Y^{(k)}_{f} S_{f}(E),
\end{equation}
where $k$ stands for $^{235}$U, $^{238}$U, $^{239}$Pu or $^{241}$Pu, $Y_{f}$ are the fission yields~\cite{Katakura2012} and the sum runs over all fission fragments. 
{In the present study, both independent and cumulative fission yields are adopted from the database of JAEA~\cite{Katakura2012} . We note that
the fission yields provided by different nuclear databases are slightly different, and that may affect the aggregate spectra.}

\begin{figure}[htb]
  \centering
  \includegraphics[width=\linewidth]{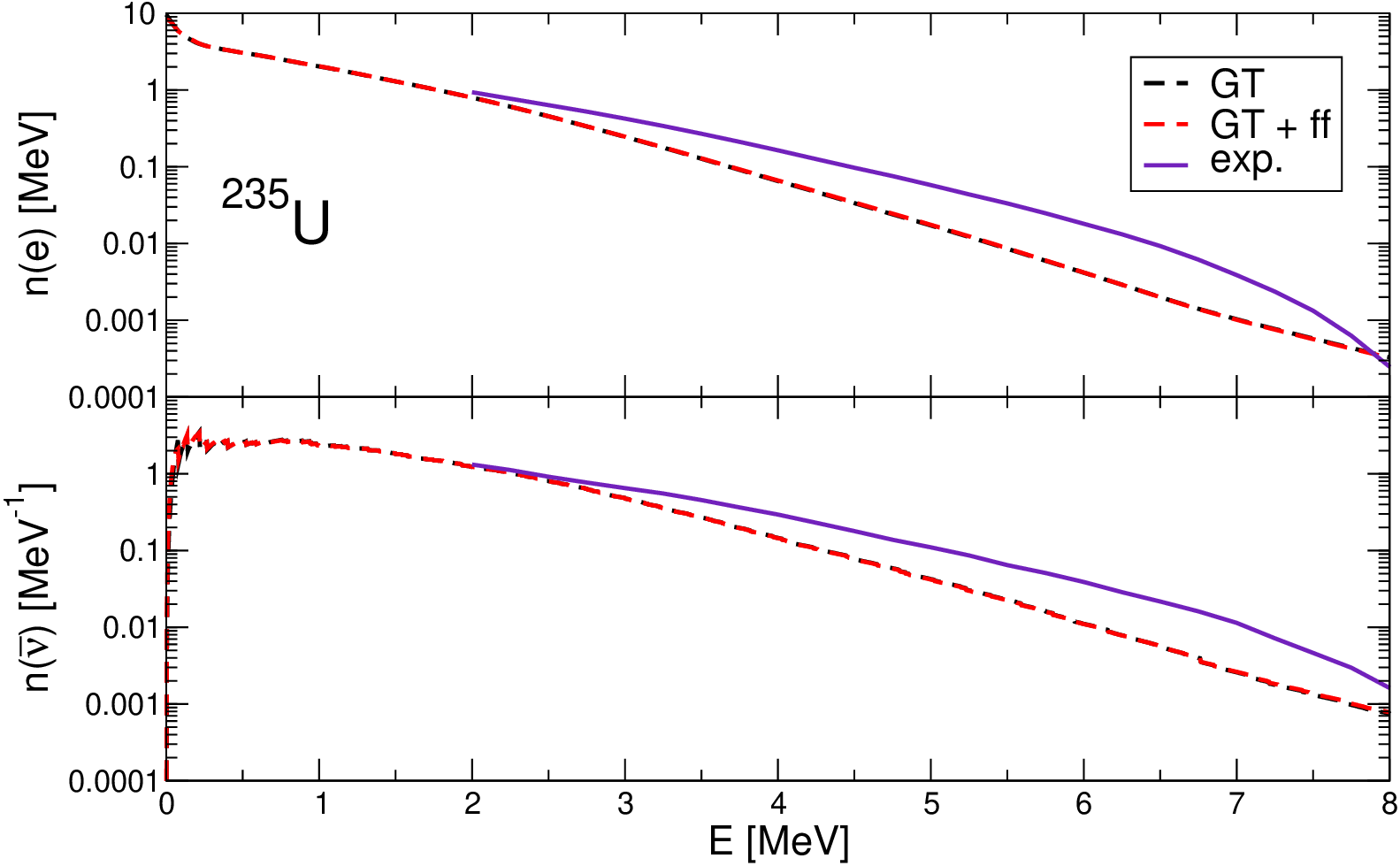}
  \caption{(color online) \label{fig:U235spectra} (top panel) The electron spectra obtained by treating all transitions as allowed (black dashed line) and electron spectra obtained by taking into account shape factors of first-forbidden transitions compared with available data. (bottom panel) Same as top panel for antineutrino spectra.}
\end{figure}

To assess the effect of first-forbidden transitions on the total lepton spectrum we perform two calculations: (i) baseline calculation where all transitions are treated as allowed, (ii) calculation where we take into account shape factors for parity changing transitions. In Fig.~\ref{fig:U235spectra} we plot the resulting electron (top panel) and antineutrino (bottom panel) spectra for $^{235}$U, where the theoretical results are denoted by full lines, and data by the dashed black line. The results for $^{238}$U, $^{239}$Pu and $^{241}$Pu are very similar and provide no additional insight.

The calculated spectra deviate significantly from the measurements, especially at high lepton energies. In particular, in the description of $\beta$ decays it is difficult to predict transitions to low-lying states in the daughter nuclei with the standard 1p-1h RPA, as it cannot describe the fragmentation and spreading of transitions. This problem can be addresses by using second RPA or particle-vibration coupling models such as in Ref. ~\cite{Robin2016}. Additionally, nuclei which contribute the most at high energies are nuclei with an odd number of nucleons which were not treated properly in the calculation of $\beta$-decay half-lives. A possible solution may be the equal filling approximation as used in Ref. \cite{Shafer2016}, where the authors observe a low-lying Gamow-Teller state (see Fig. 1. and the following discussion). In fact, a very detailed description of the structure of all the decaying nuclei is required to fully reproduce the data. This includes the properties of both the ground state and the excited states with accuracy beyond the capabilities of current models.
{We note that the calculated values for the beta decay rates may deviate from the measurements, especially for low energy transitions. However, as already 
discussed on the $\beta$-decay rates for the r-process in Ref. \cite{Marketin2016}, our agreement with measured data improves with increasing $Q$-values, and this is
the relevant aspect for the high energy part of the spectra.}

\begin{figure}[htb]
  \centering
  \includegraphics[width=\linewidth]{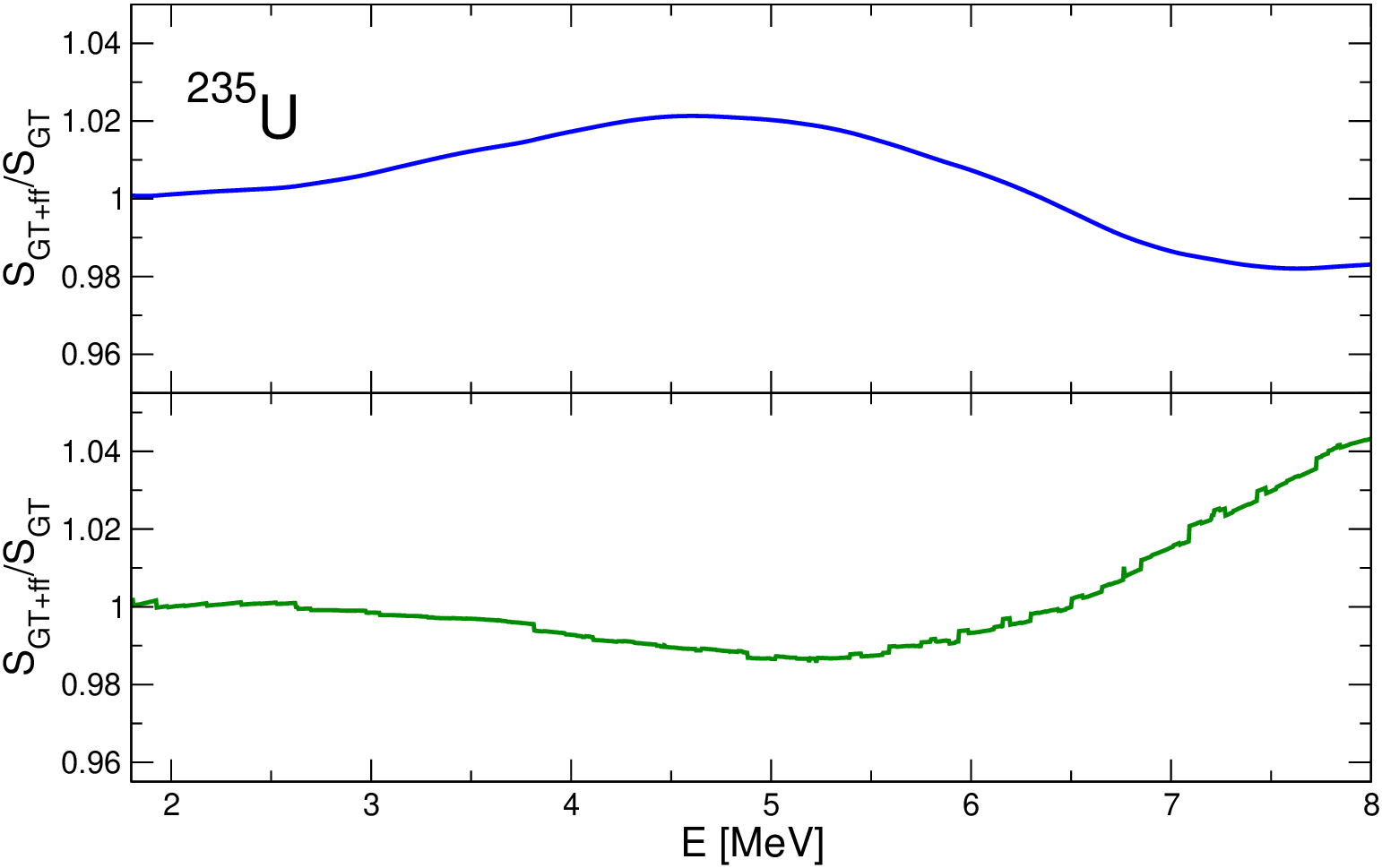}
  \caption{(color online) \label{fig:U235} (top panel) Ratio of electron spectra calculated with and without taking the first-forbidden shape factor into account. (bottom panel) Same as top panel, but for antineutrino spectra.}
\end{figure}

At the scale used in Fig.~\ref{fig:U235spectra} there is no visible difference between the two calculations, with and without taking into account the shape factor of first-forbidden transitions. This is to be expected as the magnitude of the anomaly is only 6\%, and changes of the spectrum comparable to the anomaly will not be visible. However, by examining the ratio of the spectrum obtained by taking into account the shape factor of forbidden transitions and the spectrum obtained by assuming the allowed shape for all transitions we can obtain valuable information. This ratio is shown in Fig.~\ref{fig:U235} both for the electrons, denoted by a dashed black line, and antineutrinos, denoted by the full red line. Note that the energy threshold for the detection of antineutrinos in the inverse beta decay is 1.8 MeV, thus only the results above this energy are shown.

The results indicate that, by taking into account the effect of first-forbidden transitions, the total theoretical antineutrino spectrum of $^{235}$U is lowered in the energy region from 2.5 to 6 MeV, with the largest effect being centered around 5 MeV. The magnitude of the reduction is up to 2\%, which is roughly half of the reported anomaly. The effect is also the strongest in the energy region where a shoulder was observed in all three experimental facilities (see Section 5. of Ref.~\cite{Hayes2016}). These results are systematic in that they appear in the calculation for all four contributing isotopes in the reactor core: $^{235}$U, $^{238}$U, $^{239}$Pu or $^{241}$Pu.
{Our results are comparable to previous studies, in particular, Fig.~\ref{fig:U235} displays similar effect of the forbidden transitions on the reactor antineutrino
spectra as Fig. 3. in Ref.~\cite{Hayes2014}.} In Ref.~\cite{Hayes2014}, it was found that the uncertainties introduced by forbidden transitions equal approximately 4\%, and the results of the present calculation agree with that value completely. 
For energies above 8 MeV, the majority of transitions that provide the dominant contribution to the spectra are transitions within odd-A nuclei which are very difficult to describe with the model. Additionally, the antineutrino spectra are very low at such high energies and thus we do not display the results above this energy. 

\begin{figure}[htb]
  \centering
  \includegraphics[width=\linewidth]{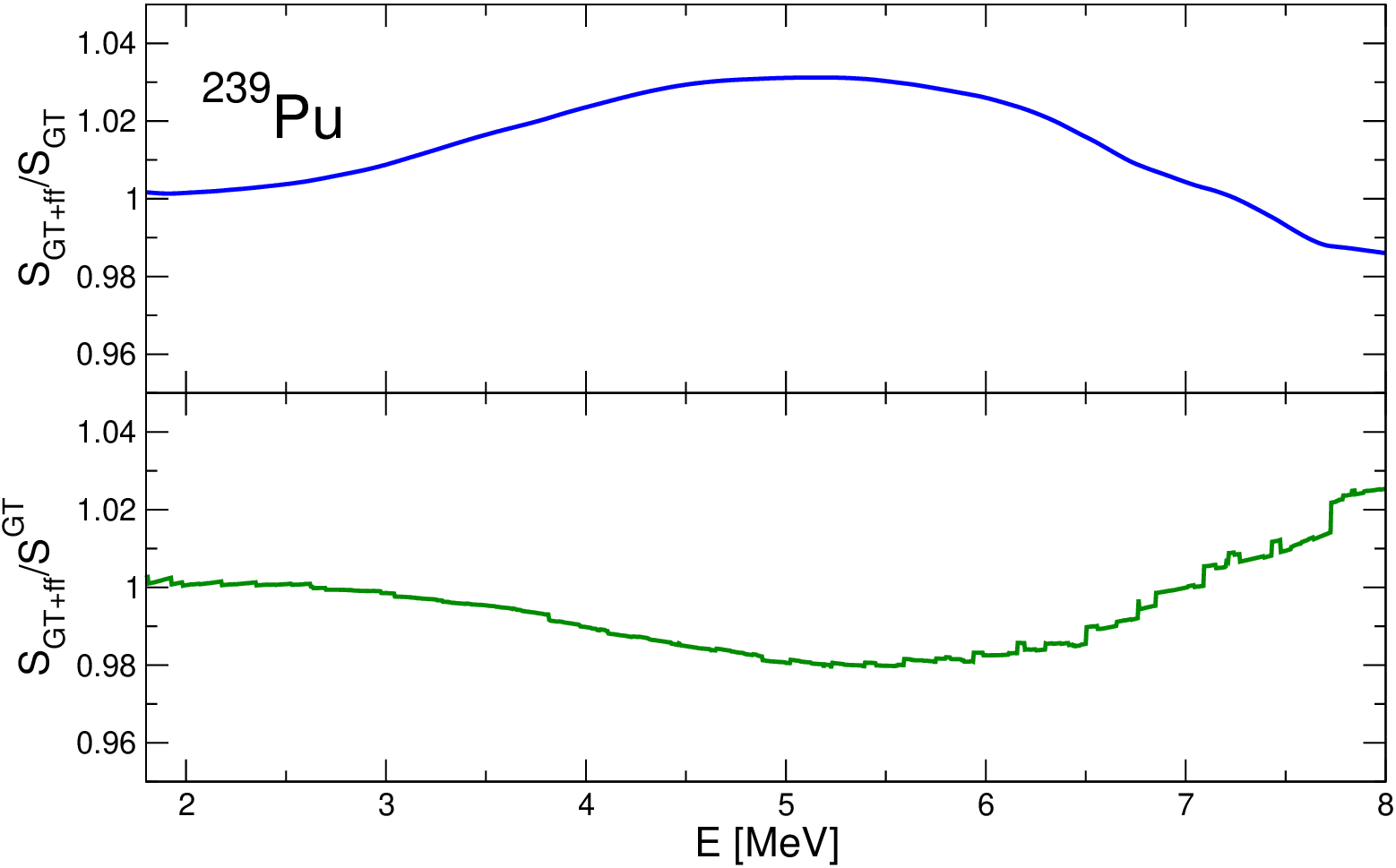}
  \caption{(color online) \label{fig:Pu239} Same as Fig. \ref{fig:U235} for the case of $^{239}$Pu.}
\end{figure}

\section{Conclusion}
In summary, we have performed the first self-consistent theoretical calculation of electron and antineutrino spectra resulting from $\beta$-decay of fission product of the main isotopes found inside a typical nuclear reactor, including forbidden transitions. In particular, the focus was on the treatment of first-forbidden transitions and their impact on the shape of the resulting lepton spectra. Having examined the three components of the first-forbidden transitions, we show that the $0^{-}$ transitions have the same shape as the allowed transitions, but the $1^{-}$ and $2^{-}$ deviate from the allowed shape and affect the total spectra significantly. By properly treating first-forbidden transitions we observe the change of the antineutrino spectra to be approximately 3\%, which is in agreement with previous studies, and is comparable to the magnitude of the anomaly itself. Therefore, proper treatment of the first-forbidden transitions is important in the study of reactor antineutrino spectra and should be taken into account in any high-precision determination of the reactor spectra.

\subsection{Acknowledgments}
This work was supported in part by the Helmholtz International Center for FAIR within the framework of the LOEWE program launched by the State of Hesse, by the 
{Deutsche Forschungsgemeinschaft (DFG, German Research Foundation) -
Projektnummer 279384907 - SFB 1245 "Nuclei: From Fundamental
Interactions to Structure and Stars"}, the IAEA Research Contract No. 18094/R0, the Croatian Science Foundation under the project Structure and Dynamics of Exotic Femtosystems (IP-2014-09-9159) and by the QuantiXLie Centre of Excellence, a project co financed by the Croatian
Government and European Union through the European Regional Development Fund, the Competitiveness and Cohesion
Operational Programme (KK.01.1.1.01). \\

\end{document}